\documentclass[pre, preprint,showpacs, showkeys,preprintnumbers,amsmath, floatfix, amssymb]{revtex4}

\usepackage{graphicx}
\usepackage{dcolumn}
\usepackage{bm}
\usepackage{float}

\begin{document}

\preprint{APS/Thorpe-Menor-1}

\title{Hierarchical Rigidity from Pair Distance Fluctuations}

\author{Scott Menor}
\affiliation{Physics Department, Arizona State University, P.O. Box 871504, Tempe, Arizona, 85287-1504}
\author{Maria Kilfoil}
\affiliation{Physics Department, McGill University
3600 Rue University, MontrŽal, Quebec, Canada H3A 2T8}
\author{M. F. Thorpe}
\email{mft@asu.edu}
\affiliation{Physics Department and Center for Biological Physics, Arizona State University, P.O. Box 871504, Tempe, Arizona, 85287-1504}

\date{\today}

\begin{abstract}
Often, experiments, observations or simulations generate large numbers of snapshots of the configurations of complex many-particle systems. It is important to find methods of extracting useful information from these ensembles of snapshots in order to document the motion as the system evolves. The most interesting information is contained in the correlated motions of individual constituents rather than in their absolute motion. We present a statistical method to identify hierarchies of correlated motions from a series of two or more snapshot configurations. This method is demonstrated in a number of systems, including freely-jointed polymer chains, hard plastic spheres, water, and proteins. These concepts are implemented as \textit{TIMME}, the Tool for Identifying Mobility in Macromolecular Ensembles. 
\end{abstract}

\pacs{61.43.Bn; 61.18.-j; 68.03.Hj}
\keywords{NMR; Collective Motions; Ensemble; Flexibility; Rigid Cluster; Co-moving Cluster; Mobility; Macromolecule; Protein; Structure; Dynamics; FIRST; TIMME; barnase; ADK}
\maketitle

\section{Introduction}

Many situations of interest generate large numbers of configurations of complex, multi-component systems. Examples include molecular dynamics simulations \cite{phillips:2005lr}, Nuclear Magnetic Resonance (NMR) structure determination experiments\cite{bycroft:1991qf, miron:2004uq}, tracking individuals in a crowd, grains of sand in a pile\cite{held:1990qy}, propagation of fire in a forest\cite{clar:1994zr, mackay:1984fr}, magnetic vortices in a superconductor\cite{field:1995fj}, rigid spheres in viscoelastic media\cite{levine:2000fk, chen:2003qy, levine:2001uq, levine:2001fj, crocker:2000kx}, cars on the freeway\cite{nagatani:1994yq, mahnke:1997vn, nagel:1995rt, kerner:1995ys}, or packets in a network\cite{ohira:1998mz, fukuda:2000ly}. These systems have significant underlying structure on various time scales and often include hierarchies of self-organized structure\cite{bak:1987kx,bak:1998uq}. Identifying that structure poses a significant challenge in how to describe and quantify the kinematics of the evolving system. 

Visually, the variations between snapshots are often characterized by direct observation and inference. Such an approach is highly subjective and qualitative. Direct observation is also severely limited because it is impossible for the human brain to simultaneously track the motion of thousands of densely packed objects. For this reason, there is a significant need for unambiguous, systematic, objective, and quantitative analysis techniques. 

For many physical quantities, the most interesting information is not the overall movement of individual objects in an ensemble of snapshots, but rather the concerted motion involving clusters of two or more objects. This correlated motion between individual component objects reflects the underlying structure of the system. Factor analytic techniques such as Principal Component Analysis (PCA) can identify correlated motion by inferring a set of basis vectors of motions that contribute most to the variation between snapshots \cite{wold:1976uq}. The PCA basis vectors can be thought of like a basis set of eigenmodes for the system, which define an easily accessible subspace. Unfortunately, PCA substitutes one complexity for another. Instead of having to consider a series of snapshots, a researcher using PCA has to consider a series of 3{\it{N}} dimensional basis vectors, where {\it{N}} is the number of objects in the system (spheres, atoms, etc.).

Experimentally, two-point microrheology uses the cross-correlated motions of pairs of beads embedded in a material and subsequently imaged via microscopy and tracked, with many beads per field of view, and relates the component along the direction of the vector separating the pair to the viscoelastic response of the material in between~\cite{crocker:2000kx,levine:2000fk}. This technique has been shown to match up with the bulk rheology in viscoelastic materials where the ensemble-averaged autocorrelated motion of individual beads differs from the bulk response~\cite{crocker:2000kx}. The disadvantages to this method are that it can only give the linear response, and that one has to average over an ensemble of pairs of beads.

Other approaches attempt to reduce the complexity to something more tractable and tangible. One example is the Hingefind algorithm\cite{wriggers:1997lr}. Hingefind assumes that large collections move as rigid bodies connected by hinges. A related approach involves `model-free' methods, which decompose into rotation vectors acting on quasi-rigid bodies\cite{hayward:1997lq, hayward:1998dq}. In model-free approaches, a system is decomposed into a collection of subdomains, where each subdomain is treated externally as a rigid body under the assumption that the low-frequency modes can be approximated as rigid-body motions. Internal motion within the quasi-rigid domains can then be treated by normal mode analysis, or by some other approximation, to reduce significantly the computational complexity. 

Hingefind and model-free approaches identify the bodies and hinges or axes of rotation, and present a simple, easily understandable picture to the researcher. Unfortunately, this simplified picture can be quite unrealistic, as real motion is often characterized by both large and small scale rearrangements, rather than by simple hinge motion between rigid bodies. 

In this paper, we present an alternative geometric approach that shares some of the strengths of PCA, Hingefind, and model-free approaches. We find that it is possible to apply a simple statistical technique to identify a hierarchy of relatively rigid or co-moving clusters, i.e. collections of objects that tend to move in concert. This hierarchy is parameterized by a single cutoff distance. Any choice of the cutoff gives a decomposition into a set of co-moving clusters. At a sufficiently high cutoff, all objects in the system are included in a single co-moving cluster, which at this level reflects rigid body motions of the entire system. As the cutoff is decreased, the single large co-moving cluster breaks up into successively smaller clusters. At a cutoff of zero, only perfectly rigid clusters, with all internal pair distances locked, are identified as co-moving. 

The collection of rigid body translations and rotations of the clusters at every level can be thought of as forming a basis for the possible motions of the system, much like in PCA. This hierarchy of motions is relatively easy to conceptualize compared with the basis vectors of a PCA analysis. Given a sufficiently large sample size, this hierarchy reflects the underlying structure of the system. In cases where the Hingefind approximation of rigid bodies joined by hinges is reasonable for a system, then at some level in the hierarchy (corresponding to a particular cutoff), the hinged rigid bodies will be identified as co-moving clusters. 

Our algorithm for identifying a hierarchy of clusters is presented in the next method section. A number of example systems are considered, in the following section, to illustrate the strengths and limitations of this approach. Finally, this algorithm is contrasted with a number of complementary analysis techniques.

\section{Method}

Consider a system of $N$ rigid objects. Assume that for any pair of objects in the system, $a$ and $b$, there exists a well-defined distance between them, $r_{a b}(t)$ at a particular time. If these objects are rigidly braced relative to each other, then $r_{a b}(t)$ will be constant over the lifetime of the system, and the standard deviation of $r_{a b}(t)$, $\sigma_{a b}$, will be zero.

\begin{eqnarray}
\sigma_{a b}^2 &=& \left<{\left(r_{a b} - \left<{r_{a b}(t)}\right>\right)^{2}}\right>\\
&=& \left<r_{a b}(t)^{2}\right> - \left<r_{ab}\right>^{2}
\end{eqnarray}

Here $\sigma_{a b}$ is a symmetric $N \times N$ matrix with $\sigma_{a b} \geq 0$ for all $a$, $b$ and $\sigma_{a a} = 0$. As the tethering between $a$ and $b$ becomes weaker, $\sigma_{a b}$ will take on some nonzero value. Completely disconnected objects will have some larger value of $\sigma_{a b}$. Given these pair distance deviations and a distance cutoff $\sigma_{cut}$, it is possible to construct a co-moving cluster decomposition. We consider any pair of objects $a$ and $b$ to be co-moving up to the cutoff $\sigma_{c}$ if $\sigma_{a b} \leq \sigma_{c}$.

In a large system, it is useful to consider an entire hierarchy of clusters. A two dimensional dilution plot is prepared by ordering the basis objects along the horizontal axis and the cutoff along the vertical axis\cite{hespenheide:2002qe}. In a dilution plot, individual clusters are shown as colored horizontal stripes at a given cutoff. Considering the entire dilution plot for a system is instructive because it can suggest which clusters are likely to be real and which occur because of statistical coincidence. These later clusters will usually be small and transient and so rather easy to identify.

Co-moving clusters are defined such that they satisfy an equivalence class relation. Given three objects $a$, $b$ and $c$, if $\sigma_{a b} < \sigma_{c}$ and $\sigma_{b c} < \sigma_{c}$, then $a$ and $c$ are considered to be members of the same cluster, even if $\sigma_{a c} > \sigma_{c}$. 
With this definition, a long, slightly flexible rod made of several basis objects will be identified as a single co-moving cluster at an appropriate choice of cutoff, as long as the motion between nearby objects does not exceed the cutoff and even if the ends of the rod individually move enough relative to one another to be placed in separate co-moving clusters in the absence of the connecting group. 

Care must be taken when applying this technique to highly constrained structures such as molecules. Because molecules are covalently bonded networks of atoms, simply treating each individual atom as an object will not produce reasonable results. The distance between a pair of covalently bonded atoms is fixed, up to thermal fluctuations, and remains fixed as the system evolves. Within a molecule, it is possible to follow a sequence of any pair of covalently bonded atoms. For this reason, the entire molecule would appear to be a co-moving cluster by the equivalence class property. The only reason the molecule might appear to decompose into components would be due to small random differences in the standard deviations of individual bond lengths, and this would only be detected with a very small cutoff.

Rather than consider individual atoms, one can consider an atom and its covalently bonded neighbors as the most basic intrinsically rigid building block. Computing the value of $\sigma_{a b}$ over all combinations of pair distances between these objects yields values that accurately reflect the actual relative motion. Although this approach is effective, it is somewhat more complicated and requires additional calculations to analyze a given system. 

More generally, the user may select appropriate intrinsically rigid objects as the basic elements. For hard shell spheres, the spheres themselves would be a natural choice. For bonded atoms in molecules, the natural choice would be the object defined by an atom and its bonds but excluding its bonded neighbors. 

The above concepts are implemented as \textit{TIMME}, a Tool for Identifying Mobility in Macromolecular Ensembles. An outline of the \textit{TIMME} algorithm is given by the following list:

\begin{enumerate}
\item Select an appropriate intrinsically rigid object to act as the fundamental unit for the analysis of the system.
\item \label{timmeAlgorithmIterateStep}Iterate through each pair of objects $a$ and $b$ in the system.
\item Compute $\sigma_{a b}$ for each pair of objects.
\item Sort the values of $\sigma_{a b}$ in increasing order, and join together the clusters containing sites $a$ and $b$ into a composite co-moving cluster. This has the effect of including all sites with $\sigma_{i j} \leq \sigma_{a b} = \sigma_{c}$
\item Accumulate statistics for the co-moving clusters at each $\sigma_{a b}$. In particular, compute for each $\sigma_{c}$ the fraction of sites contained in co-moving clusters containing 10 or more sites. 
\item Assign each site a unique label. 
\end{enumerate}

For a system of $N$-obects, step \ref{timmeAlgorithmIterateStep} scales as $O(N^{2})$. There are several obvious simplifications to the \textit{TIMME} algorithm. In particular, considering only spatially adjacent pairs rather than all pairs reduces this step to one of $O(N)$, rather than $O(N^2)$. 

\section{Examples}

The \textit{TIMME} algorithm makes no assumptions about the source of input data. It can be applied equally well to a set of snapshot coordinates from NMR structure determination experiments, confocal microscopy direct visualization experiments, molecular dynamics simulations, people within a crowd, or cars on a road. Although all of the examples presented here are confined to three dimensional coordinates, this approach can also be used to find clustering hierarchies for arbitrary data embedded in any high-dimensinal space. 

Several example systems follow. Each was selected to demonstrate specific properties of the \textit{TIMME} algorithm and features of the results it generates. The examples covered in the next sections are: {\it{A.}} the freely jointed polymer chain; {\it{B.}} hard spheres; {\it{C.}} water; and {\it{D.}} proteins.

\subsection{Freely-Jointed Polymer Chain}

As a model of a linear polymer, we consider a freely-jointed chain in 3-dimensions \cite{edwards:1972dk}. Each monomer subunit is modeled as a rod of length $a$. Linked pairs of monomers are linked by ball-and-socket hinges, all possible angles are sampled with equal probability. Overlaps between individual chain elements are allowed. 

While it is tempting to consider the joints as the fundamental objects for this system, such a choice is poor because any pair of adjacent joints will always be separated by a constant length. If the joints were used as the fundamental objects, then a \textit{TIMME} analysis would conclude that the entire polymer chain was a single co-moving cluster for any choice of cutoff, $\sigma_{c}$. A better choice is to use the links (bonds) between joints (atoms) as fundamental objects for the freely-jointed chain. 

\begin{figure}[H]
\begin{center}
\includegraphics[width=.7\columnwidth]{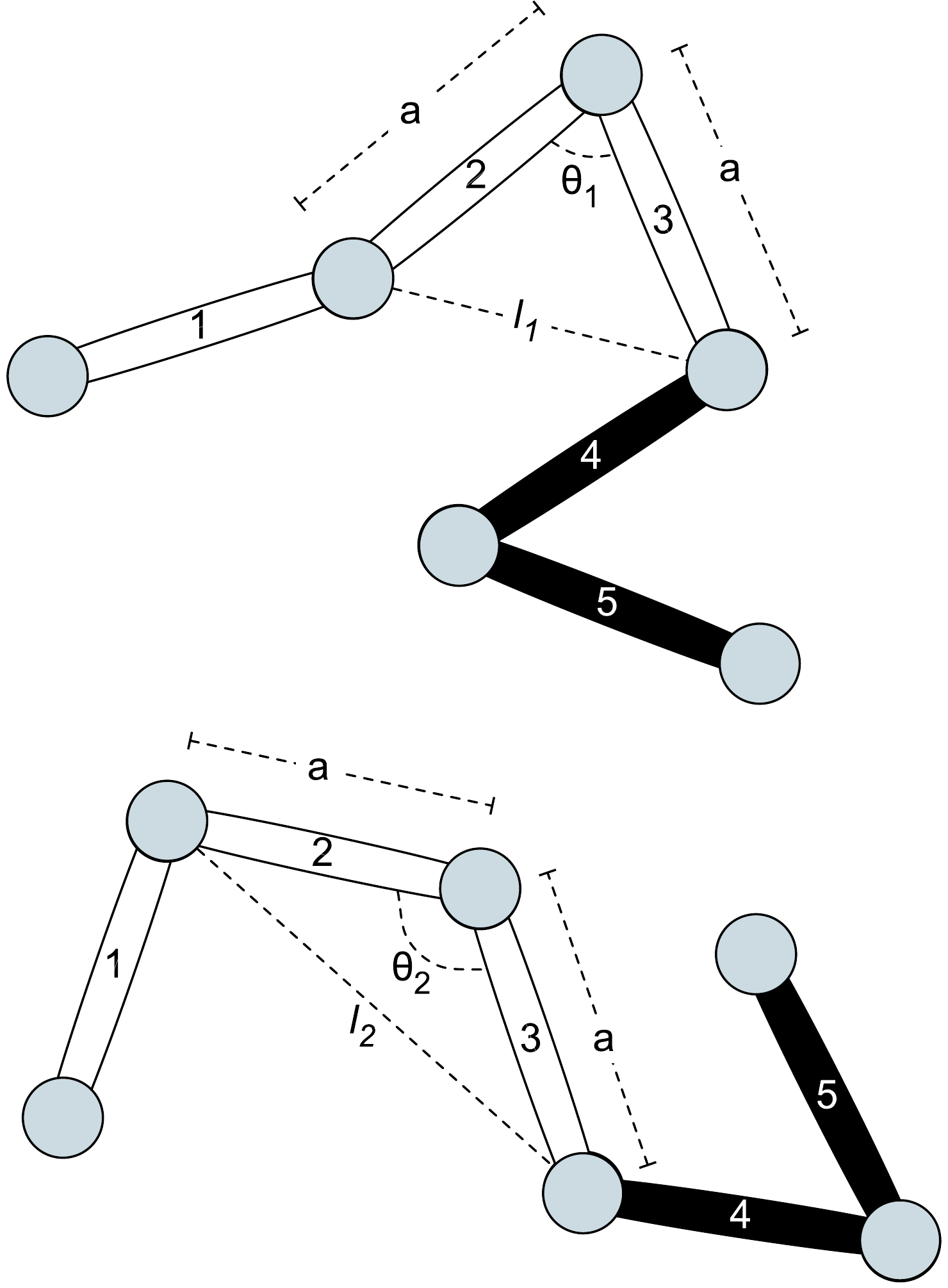} 
\caption{Two realizations of a freely-jointed chains with 6 joints and 5 links. Each link has length $a$. The angle between links $2$, and $3$ is $\theta_{1}$ in the first realization and $\theta_{2}$ in the second realization. The corresponding pair distance between neighboring links ($2$ and $4$, in this case) is given by $l_{1}$ in the first realization and $l_{2}$ in the second. The relationship between links $4$, and $5$ remains constant (rigid) between the two structures, as indicated by the solids links, while the rest of the structure is more flexible}
\label{fig:freelyJointedChains}
\end{center}
\end{figure}

The separation between the two next-nearest neighbor joints, as shown in Fig. \ref{fig:freelyJointedChains}, is

\begin{equation}
l = a \sqrt{2(1-cos(\theta))}\,.
\end{equation}
The probability $p(l)$ of measuring a value $l$, is
\begin{eqnarray}
p(l)&=& \frac{1}{2} \int_{0}^{\pi}d\theta sin(\theta) \delta\left(l- a \sqrt{2(1-cos(\theta))}\right) \label{eq:probabilityForFreelyJointedChain}\\
&=& \frac{l}{2 a^{2}}\,.
\end{eqnarray}
For two random snapshot realizations, the standard deviation of the length, $l$, from its mean, $\left<l\right>$, is  
\begin{eqnarray}
\sigma^{2}&=&\frac{1}{2}\sum_{s=1}^{2}\left(l_{s} - \left<l\right>\right)^{2}\\
&=&\frac{1}{2}\sum_{s=1}^{2}\left(l_{s} - \frac{1}{2} \sum_{t=1}^{2}l_{t}\right)^{2}\\
&=&\frac{1}{4}\left(l_{1}-l_{2}\right)^{2}\,.
\end{eqnarray}
It follows that the probability density of $\sigma$, $p(\sigma)$, is 
\begin{eqnarray}
p(\sigma)&=& \int_{0}^{2 a}dl_{1} p(l_{1}) \int_{0}^{2 a} dl_{2} p(l_{2}) \delta \left( \sigma - \frac{1}{2}\left|l_{1}-l_{2}\right|\right) \\
&=& 2 \int_{0}^{2 a} \frac{dl_{1} l_{1}}{2 a^{2}} \int_{0}^{l_{1}} \frac{dl_{2} l_{2}}{2 a ^{2}} \delta \left( \sigma - \frac{1}{2}\left(l_{1}-l_{2}\right)\right) \\
&=& \frac{4(a-\sigma)^{2}(2 a+\sigma)}{3 a^{4}}\label{eq:pOfSigma2} 
\end{eqnarray}
for $0 < \sigma < a$, which gives 
\begin{eqnarray}
p_{c} = p(\sigma < \sigma_{c}) &=& \int_{0}^{\sigma_{c}} d\sigma p(\sigma) \\
&=& \int_{0}^{\sigma_{c}} d\sigma  \frac{4(a-\sigma)^{2}(2 a+\sigma)}{3 a^{4}}\\ 
&=&\frac{\sigma_{c}}{3 a^{4}} \left(8 a^{3} - 6 a^{2} \sigma_{c}+\sigma_{c}^{3}\right)\,.
\end{eqnarray}

\begin{figure}[H]
\centering
\begin{tabular}{ccc}
$r = 1$ &
\begin{minipage}{2.2in}
\centering
\includegraphics[width=2in]{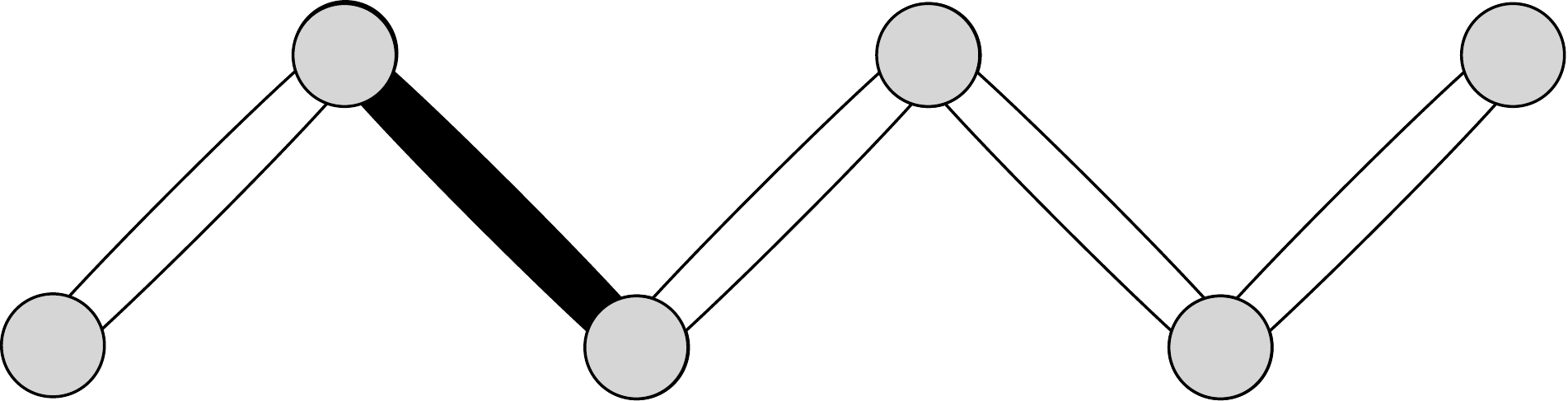} 
\end{minipage} & $\left(1-p_{c}\right)^{2}$ \\
\\$r = 2$ &
\begin{minipage}{2.2in}
\centering
\includegraphics[width=2in]{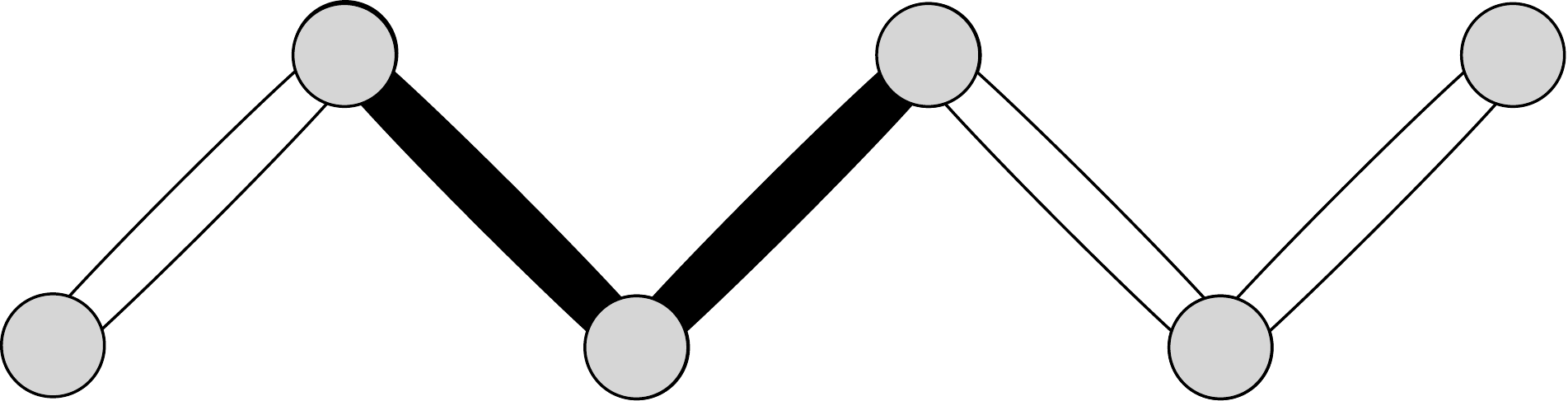} 
\end{minipage}& $2 \left(1-p_{c}\right)^{2} p_{c}$\\
\\$r = 3$ &
\begin{minipage}{2.2in}
\centering
\includegraphics[width=2in]{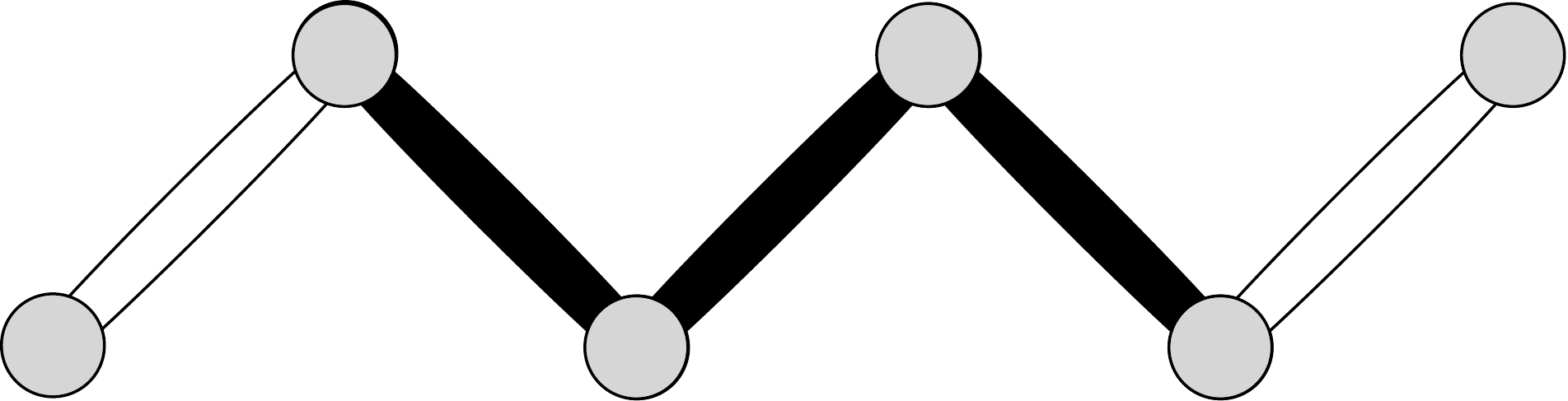} 
\end{minipage}& $3 \left(1-p_{c}\right)^{2} p_{c}^{2}$ \\
\end{tabular}
\caption{Diagrams of clusters of links in freely jointed chains. Each vertex represents a joint. Each line segment represents a linkage. Dark segments represent clusters of linkages. In general, it can be shown that the fraction of sites in a cluster of size $r$, $P(r)$ is $r (1-p_{c})^{2} p_{c}^{r-1}$, as in Eq. (\ref{eq:pOfR}).\label{fig:freelyJoinedChainDiagram}}
\end{figure}

At some cutoff, $\sigma_{c}$, the probability of two adjacent linkages being joined is $p_{c} = p(\sigma < \sigma_{c})$. The probability of two adjacent linkages being not joined is $1 - p_{c}$. For a sufficiently long chain, any cluster must begin and end with disconnected links with a corresponding probability of $\left(1-p_{c}\right)^{2}$. Every vertex joining two links within a cluster has probability $p_{c}$. There are $r$ links in a cluster of size $r$, as shown for $r$ from 1 to 3 in Fig. \ref{fig:freelyJoinedChainDiagram}. Therefore, the probability $ P(r)$ of a site being in a cluster of $r$ links is
\begin{equation}
P(r) = r (1-p_{c})^{2} p_{c}^{r-1}
\label{eq:pOfR}
\end{equation}
The expectation value or $r$, $\left<r\right>$, is given by 
\begin{eqnarray}
\left<r\right> &=& \sum_{r=1}^{\infty} r P(r)\\
&=& \sum_{r=1}^{\infty} r^{2} (1-p_{c})^{2} p_{c}^{r-1} \\
&=&\frac{1+p_{c}}{1-p_{c}}\,.
\end{eqnarray}
For small values of $p_{c}$ (corresponding to a small cutoff, $\sigma_{c}$), $\left<r\right> = 1$, corresponding to single-site clusters. $\left<r\right>$ increases monotonically with $p_{c}$, and diverges as $p_{c}$ approaches $1$, where all sites will be members of a single cluster. 

\begin{figure}[H]
\begin{center}
\includegraphics[width=\columnwidth]{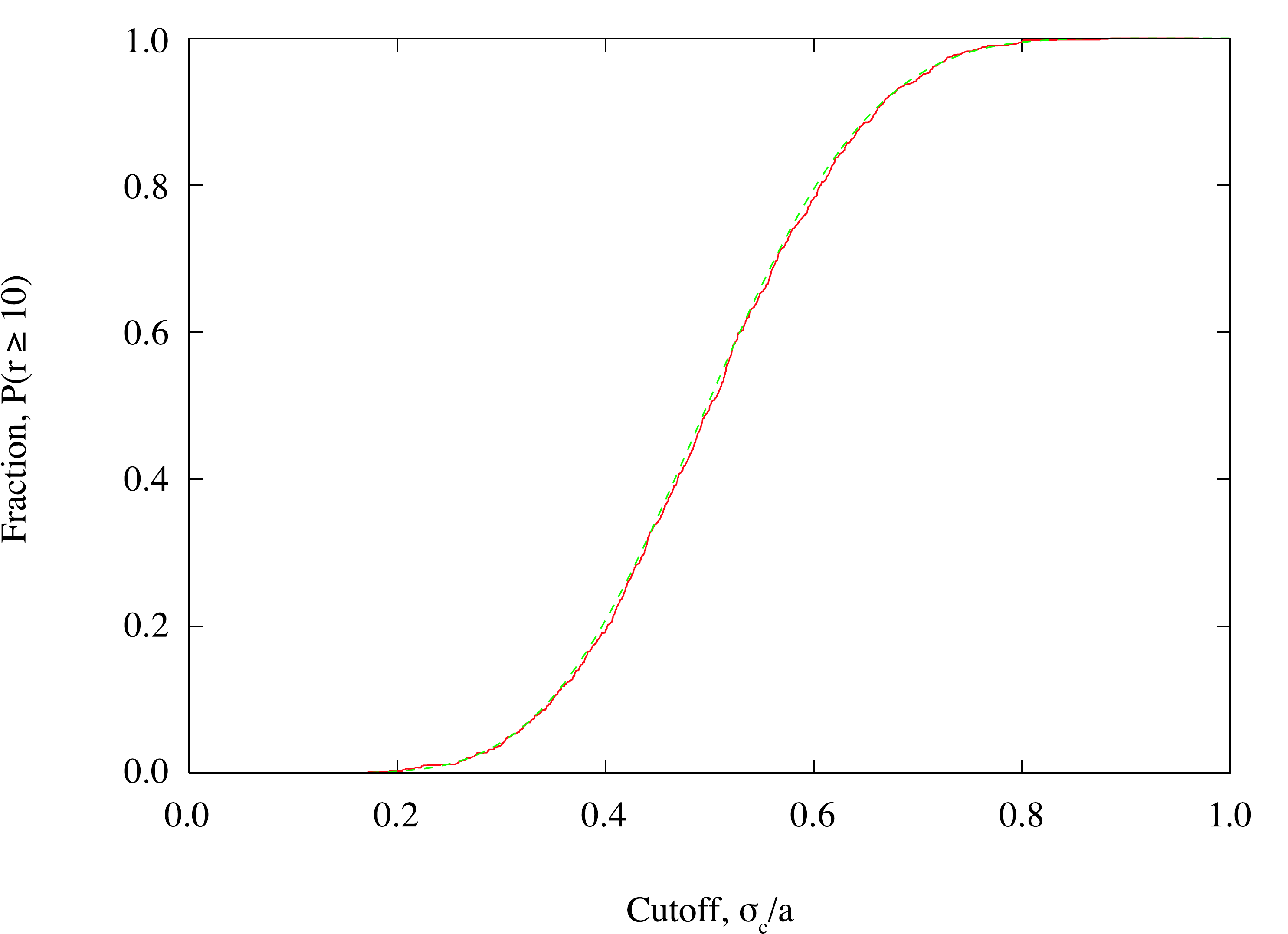}
\caption{The fraction of sites in clusters of 10 or more plotted versus $\sigma_{c} / a$, where $a$ is the length of each chain link, from theory (dashed) in Eq. (\ref{eq:freelyJointedChainSigmoid}), and from a simulated 10,000 link long freely-jointed chain (solid).}
\label{fig:freelyJointedChainSigmoid}
\end{center}
\end{figure}

These ideas can be extended to $N>2$ conformations, but we have been unable to find a general closed form expression for $p(\sigma)$, except for $N=2$ as given in Eq. (\ref{eq:pOfSigma2}). However, it is possible to calculate the first few moments \cite{chubynsky:2007fn} of $p(\sigma)$ defined via the second moment, 
\begin{equation}
\left<\sigma^{2}\right> = \frac{2}{9}a^{2}\left(1-\frac{1}{N}\right)\label{eq:secondMomentOfSigmaForNframes}\,.
\end{equation}
and the variance
\begin{equation}
\left<\sigma^{4}\right>-\left<\sigma^{2}\right>^{2} = \frac{28}{405}\frac{a^{4}}{N}\left(1-\frac{1}{N}\right)^{2} + \frac{8}{81}\frac{a^{4}}{N^{2}}\left(1-\frac{1}{N}\right)\label{eq:varienceOfSigmaForNframes}\,.
\end{equation}
which reduce to the results that can be obtained directly by integration from Eq. (\ref{eq:pOfSigma2}) when $N=2$. Similar results can be obtained for \textit{any} distribution that would replace the freely rotating chain given in Eq. (\ref{eq:probabilityForFreelyJointedChain}). These results have some important implications that are independent of this particular model. The first is that the second moment of the distribution is only very weakly dependent on the number of conformations $N$ used, which is important as it means that the sampling intervals are not important as long as the cluster stays together and remains intact. This is a quite general result as is intuitively clear. The second is that the variance, Eq. (\ref{eq:varienceOfSigmaForNframes}), goes to zero as $N \rightarrow \infty$, meaning that the distribution, $p(\sigma)$, becomes a delta function, and that all the clusters are of size zero for $\sigma < \sigma_{c}$ and there is a single large cluster for $\sigma > \sigma_{c}$, where for the freely rotating chain, $\sigma_{c} / a = \sqrt{2/9} = 0.471$, from Eq. (\ref{eq:secondMomentOfSigmaForNframes}). This means that Fig. \ref{fig:freelyJointedChainSigmoid} sharpens up to a step function at $\sigma_{c}/a = 0.471$, as the number of conformations of the freely rotating chain increases to infinity, which we have confirmed with simulations analyzed with \textit{TIMME}. This result is special to the case where all elements are equivalent, and would not be true for the models considered in the next sections of this paper, which is of most interest. Quite generally, there is little dependence on the number of conformations, $N$, used in the sampling, provided, of course, that clusters do actually stay together at a particular cutoff, $\sigma_{c}$. 

\begin{figure}[H]
\begin{center}
\includegraphics[width=\columnwidth]{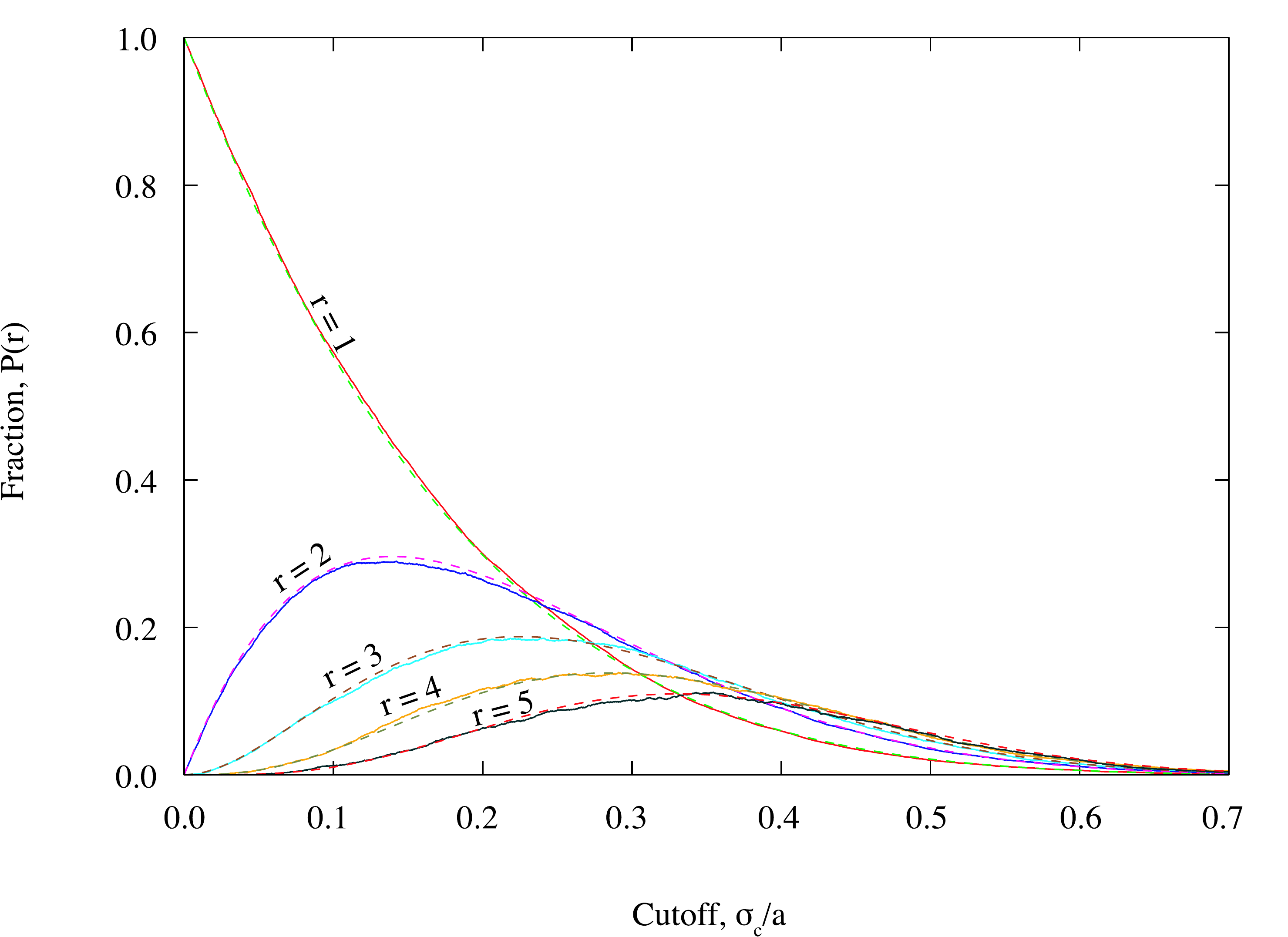}
\caption{The fraction of sites in a cluster of size $r$ ranging from 1 to 5 versus $\sigma_{c}/a$, where $a$ is the length of each chain link, from theory (dashed; see Eq. (\ref{eq:pOfR})) and from a simulated 10,000 link long freely-jointed chain (solid). }
\label{fig:freelyJointedChain}
\end{center}
\end{figure}

Two independent 10,000-link freely-jointed chain realizations were created by randomly assigning each link an orientation from a uniform distribution over the surface of a sphere of radius $a$. Only nearest neighbors were considered during the \textit{TIMME} analysis. There was good correspondence between the model, Eq. (\ref{eq:pOfR}) and the simulation, as can be seen in Fig. \ref{fig:freelyJointedChain}
It can be useful to lump the largest clusters together via
\begin{eqnarray}
P(r \geq R) &=& \sum_{r=R}^{\infty} P(r) \\
&=& 1 - \sum_{r=0}^{R-1} P(r)\,. \label{eq:freelyJointedChainSigmoid}
\end{eqnarray}
where $P(r \geq R)$ represents the fraction of sites contained in a cluster of size greater than or equal to $R$. When all associations are by chance alone, this fraction typically takes a sigmmoidal form, as in Fig. \ref{fig:freelyJointedChainSigmoid}.

\subsection{Hard Spheres with Added Weak Attraction}

An interesting mesoscopic physical model consists of small polymethyl methacrylate (PMMA) spheres contained between two glass microscope slides and driven by an added interparticle attractive interaction of strength at contact of order $k_BT$. This system can serve as a model for metallic glasses\cite{prasad:200ec, segre:2001bv} and other random sphere packs\cite{gao:2004pb, dinsmore:2002lh, dinsmore:2001ff}.

The influence of gravity is minimized by immersing the spheres in a density-matched fluid. The full three-dimensional coordinates of each sphere of a collection of hundreds to thousands of spheres within an imaging volume are observed using confocal microscopy in conjunction with shape recognition algorithms. The trajectory of individual spheres through time can then be tracked by acquiring a succession of closely spaced snapshots at equally-spaced time intervals chosen to access the relevant timescales for capturing the motions \cite{prasad:200ec, segre:2001bv, dinsmore:2001ff, gao:2007lr}. 

Because they are weakly bonded, such systems of nearly-identical spheres organize into relatively rigid network structures and relatively free mobile clusters of one or more spheres, on timescales that are not too long. At long times, a bimodal distribution of motions was detected in the experimental analysis, corresponding to two populations of relatively jammed and relatively free spheres~\cite{gao:2007lr}. It is difficult to visually identify and distinguish between these two types of structures in a homogenous, packed population. These transiently jammed features are readily apparent in a \textit{TIMME} analysis. 

\begin{figure}[H]
\begin{center}
\includegraphics[width=\columnwidth]{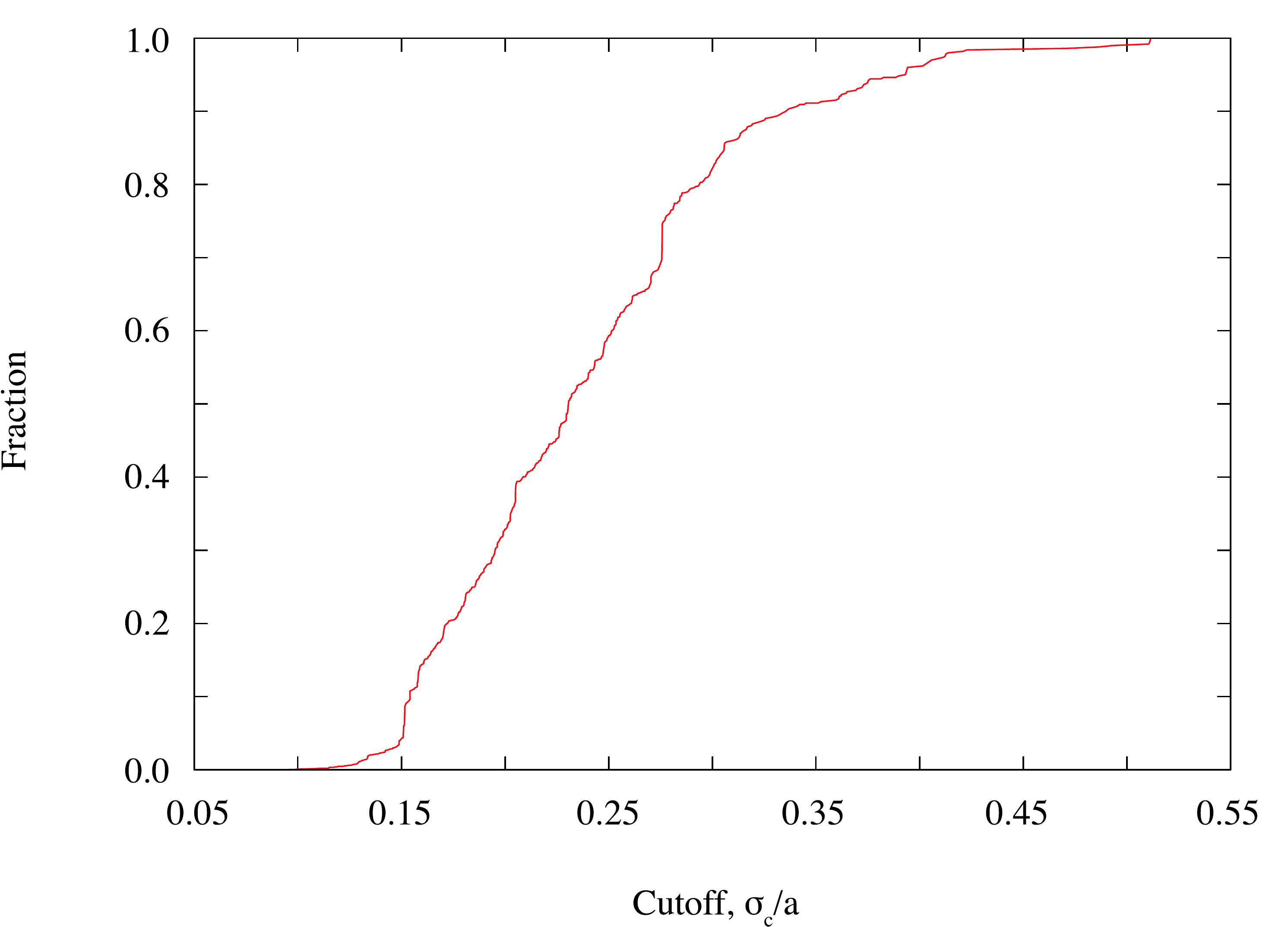}
\caption{The fraction of PMMA spheres contained within co-moving clusters containing 10 or more spheres, as a function of scaled cutoff ($\sigma_{c} / a$, where $a = $ 1.3$\mu m$, the diameter of the spheres). }
\label{fig:PMMASpheresLargestClusterStatistics}
\end{center}
\end{figure}

\textit{TIMME} analysis was performed on a data set consisting of an ensemble of 46 snapshots of 991 PMMA spheres \cite{gao:2004pb, kilfoil:2003fu}. To ensure that the identified clusters are continuous in at least one snapshot, only pairs of spheres that were in contact in one or more frames of the confocal microscopy data were considered in the analysis. As in the freely-jointed chain, the fraction of spheres in clusters of 10 or more spheres grows roughly sigmoidally with cutoff, as seen in Fig. \ref{fig:PMMASpheresLargestClusterStatistics}. 

\begin{figure}[H]
\begin{center}
\includegraphics[width=\columnwidth]{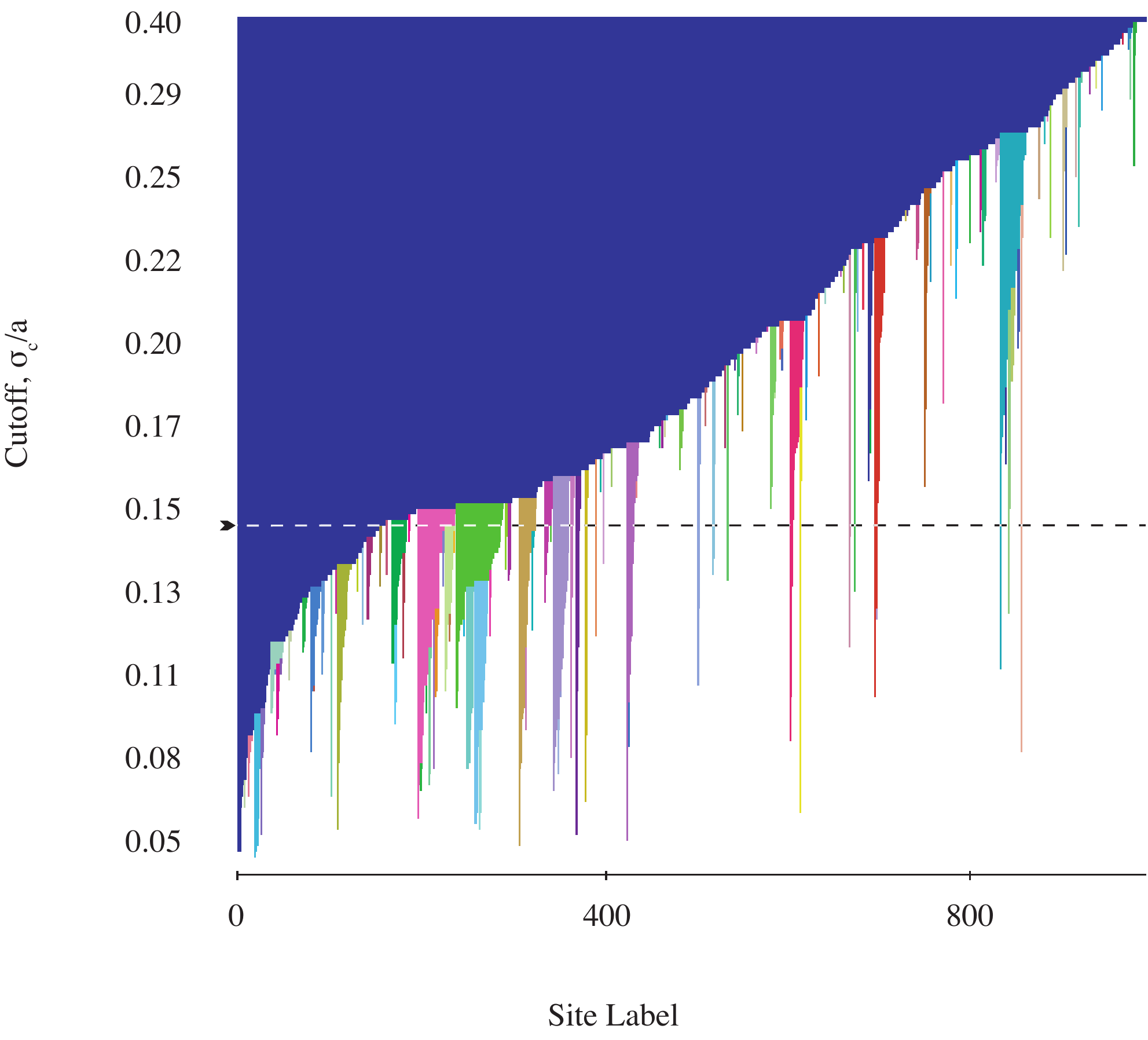}
\caption{Dilution of the clusters identified by \textit{TIMME} for 46 snapshots of a system comprising 991 PMMA spheres observed by confocal microscopy. Analysis was performed using a sparse matrix of $\sigma_{a b}$ treating only physically adjacent pairs as nonzero. Co-moving clusters consisting of 3 or more spheres are shown in various colors, with the largest in dark-blue. Clusters containing fewer than 3 spheres are suppressed in this plot for clarity. The dashed line shows the cutoff used in Fig. \ref{fig:PMMASpheresKissingNetwork}} \label{fig:PMMASpheresDilutionPlot}
\end{center}
\end{figure}
\begin{figure}[H]
\begin{center}
\includegraphics[width=\columnwidth]{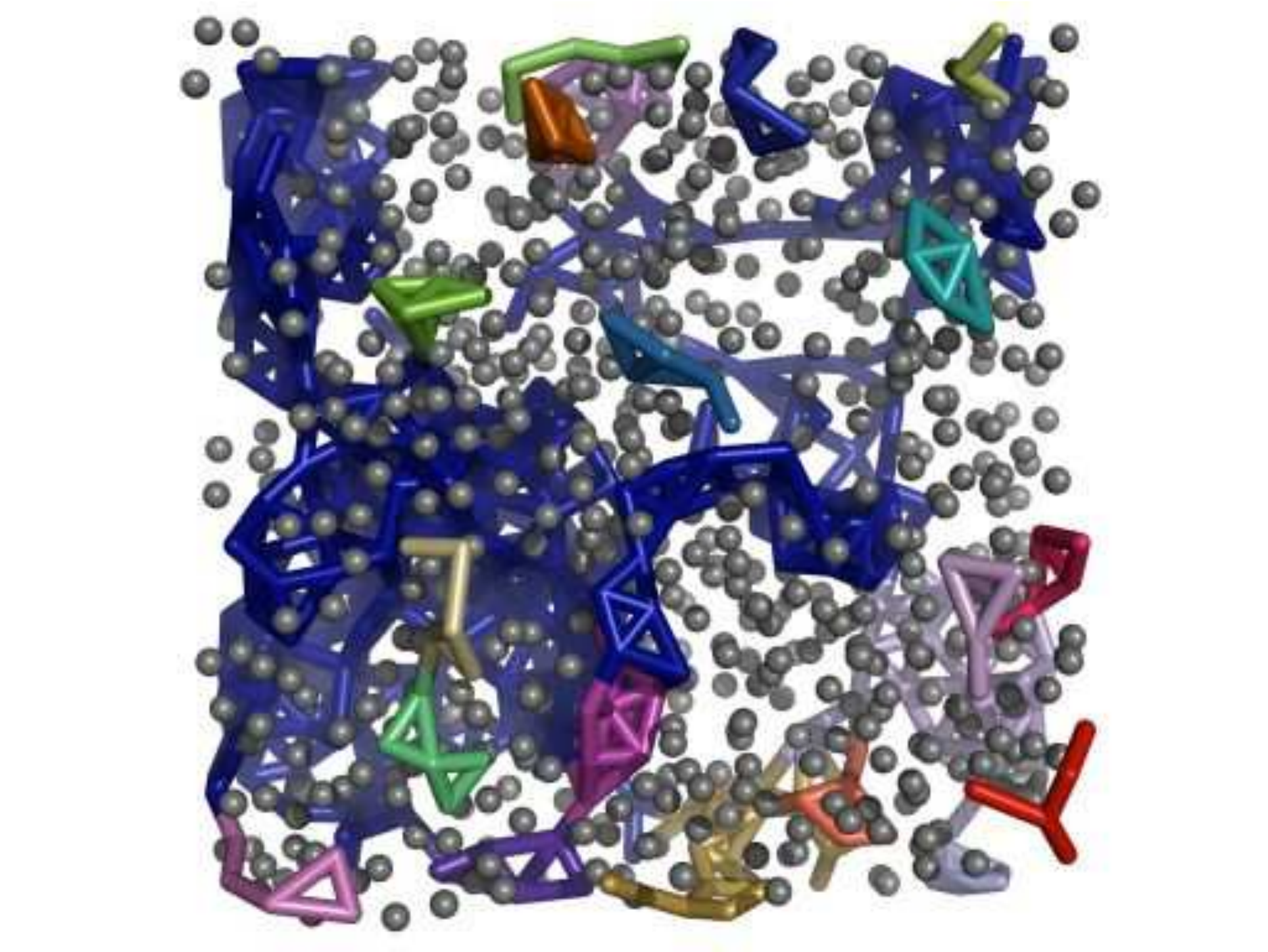}
\caption{Top-down view of a single representative snapshot from a collection of 46 snapshots of a system of 991 PMMA spheres, showing a kissing network. Co-moving clusters were computed at a cutoff of $\sigma_{c} / a = $ 0.146 (the dashed line in Fig. \ref{fig:PMMASpheresDilutionPlot}). Clusters consisting of 3 or more spheres are shown as thick sticks between individual spheres in various colors, with the largest, or the `co-moving core', in dark-blue. Spheres that are not members of larger co-moving clusters are small grey balls shown here at approximately 1/2 their actual size to allow the viewer to see past them.}
\label{fig:PMMASpheresKissingNetwork}
\end{center}
\end{figure}

In many systems, such as these PMMA spheres, there is no natural ordering for the basis objects. In such systems, it is convenient to define a 'site label' based on \textit{TIMME} dilution. This site label is chosen such that at all cutoffs, any particular co-moving cluster is continuous. Additionally, as the cutoff decreases and clusters bifurcate, sites in the largest cluster are assigned the smaller site labels so that in dilution plots the larger clusters will tend to segregate to the left and smaller clusters to the right. Although this condition is imposed at each breaking point in the dilution, it is possible that a cluster that was assigned smaller site labels will later break up and become smaller than a cluster with larger site labels. 

These site labels are used in the dilution plot shown in Fig. \ref{fig:PMMASpheresDilutionPlot}. An inflection in the growth of the core at a cutoff of approximately $\sigma_{c} / a = $ 0.147 reflects a transition between detecting physically jammed clusters and random clusters. A representative snapshot realization is shown in Fig. \ref{fig:PMMASpheresKissingNetwork} at a cutoff just below the inflection point. 

\subsection{Water}

A simulated droplet of 1000 explicit water molecules was equilibrated at room temperature for 0.8 ns. This droplet was then subjected to 0.2 ns molecular dynamics simulation using \textit{NAMD}\cite{phillips:2005lr} at standard temperature (273.15 Kelvin) and pressure (1 atmosphere). 

In real water, individual molecules can auto-ionize into hydroxide and hydronium ions. Such behavior is not present this molecular dynamics run. Thus, the simulated drop of pure water consisted of identical molecules. Like the system of PMMA spheres, there is the potential for geometric jamming creating another source for co-moving structures on some timescale. In addition to the hard-shell repulsion, water molecules can form transient hydrogen bonds. These bonds allow groups of two or more water molecules to form short-lived structures such as tetrahedra, chains, and rings\cite{wernet:2004jw, mamonova:2005fk, head-gordon:2006tw}.

\begin{figure}[H]
\begin{center}
\includegraphics[width=\columnwidth]{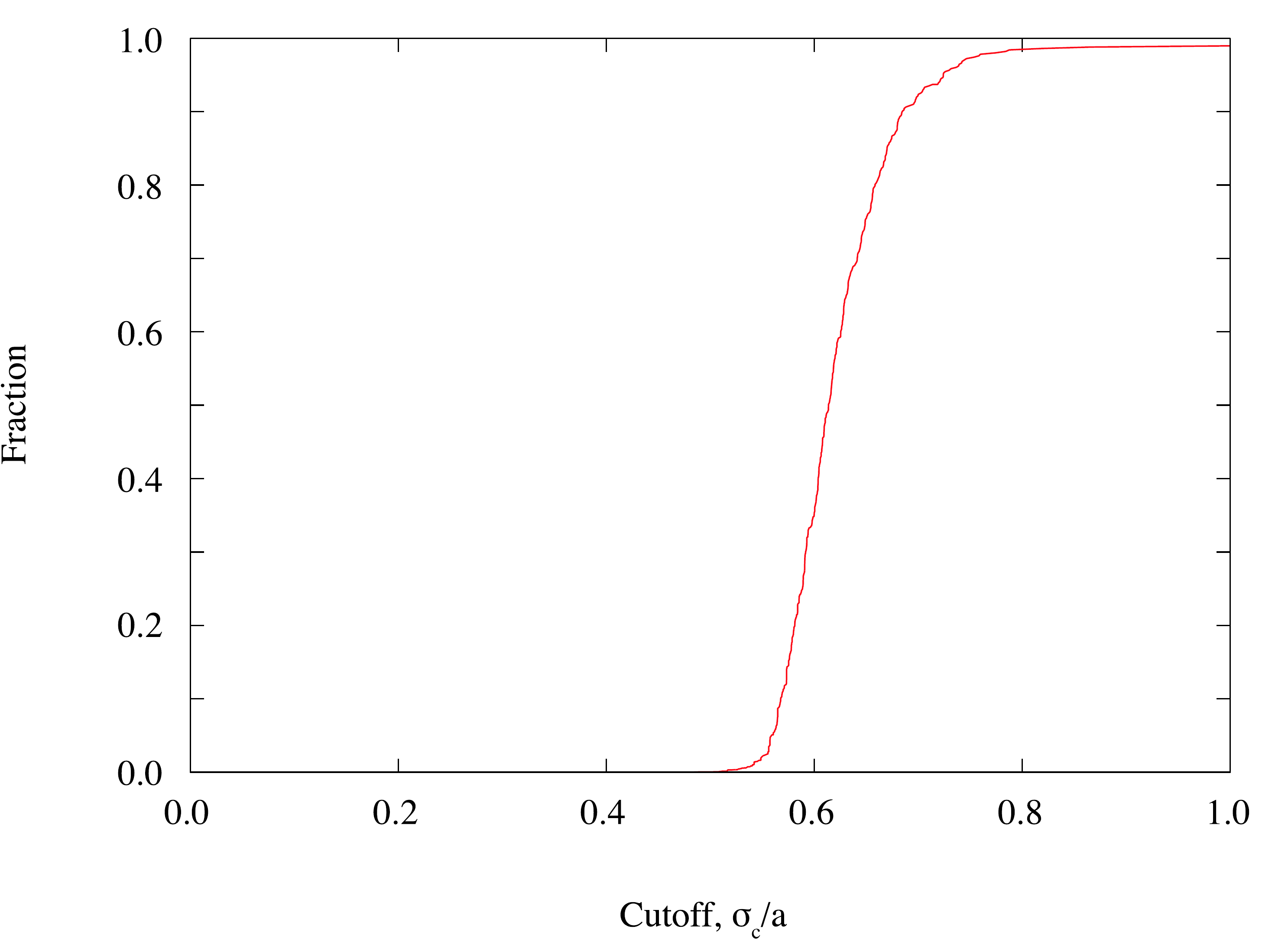}
\caption{The fraction of atoms in co-moving clusters of 10 or more water molecules, as identified by $TIMME$, in a 0.2 ns \textit{NAMD}\cite{phillips:2005lr} molecular dynamics simulation of 1000 water molecules in vacuum as a function of cutoff, $\sigma_{c}$, normalized by the oxygen-hydrogen bond length ($a = $ 0.96\AA). }
\label{fig:waterDropLargestClusterStatistics}
\end{center}
\end{figure}

\begin{figure}[H]
\begin{center}
\includegraphics[width=\columnwidth]{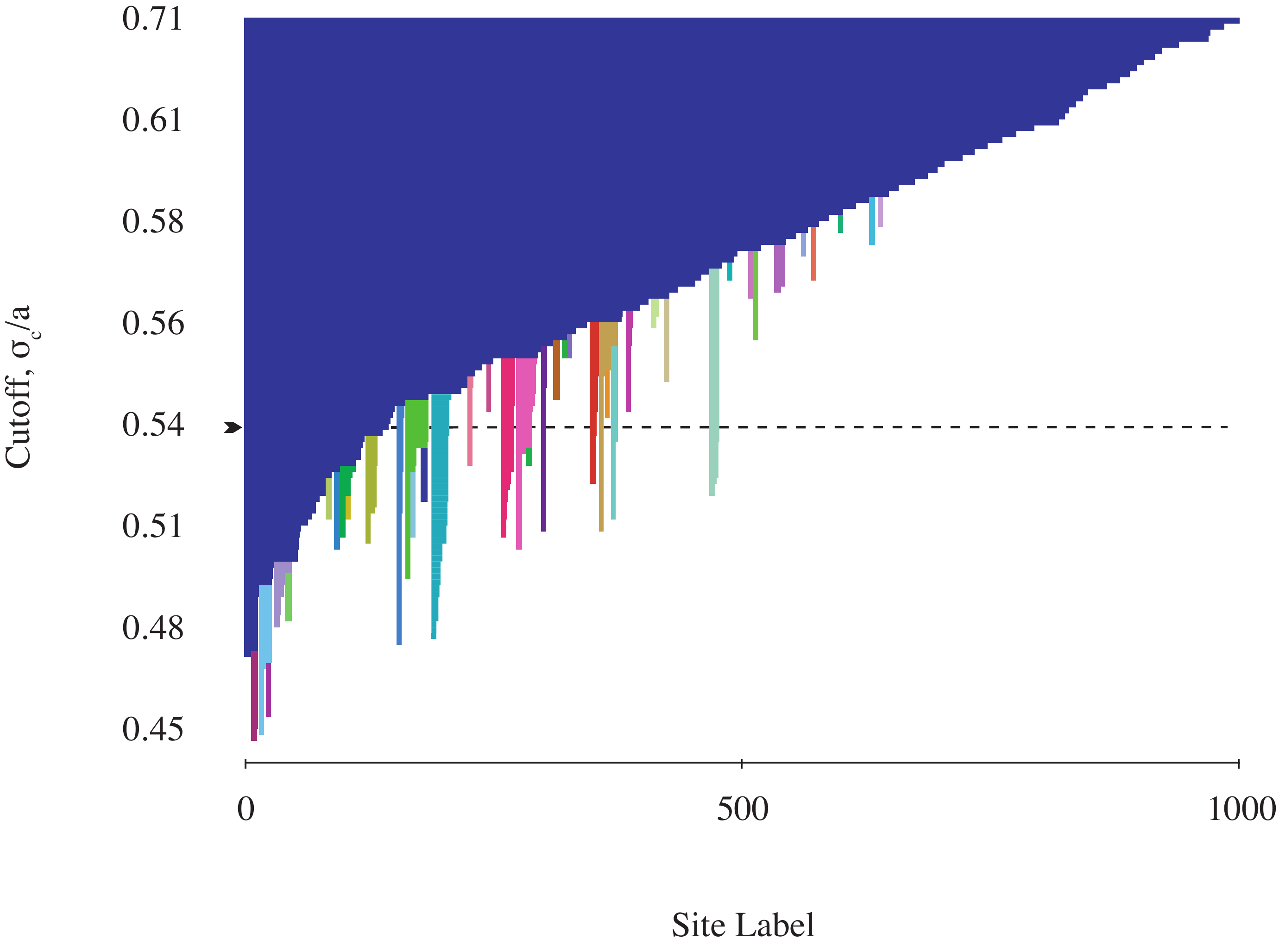}
\caption{Dilution plot of the clusters identified by \textit{TIMME} for a 0.2 ns MD trajectory of a liquid water drop consisting of 1000 molecules at standard temperature and pressure. The dashed line shows the cutoff used in Fig. \ref{fig:waterDropClusters}} 
\label{fig:waterDropDilutionPlot}
\end{center}
\end{figure}

\begin{figure}[H]
\begin{center}
\includegraphics[width=\columnwidth]{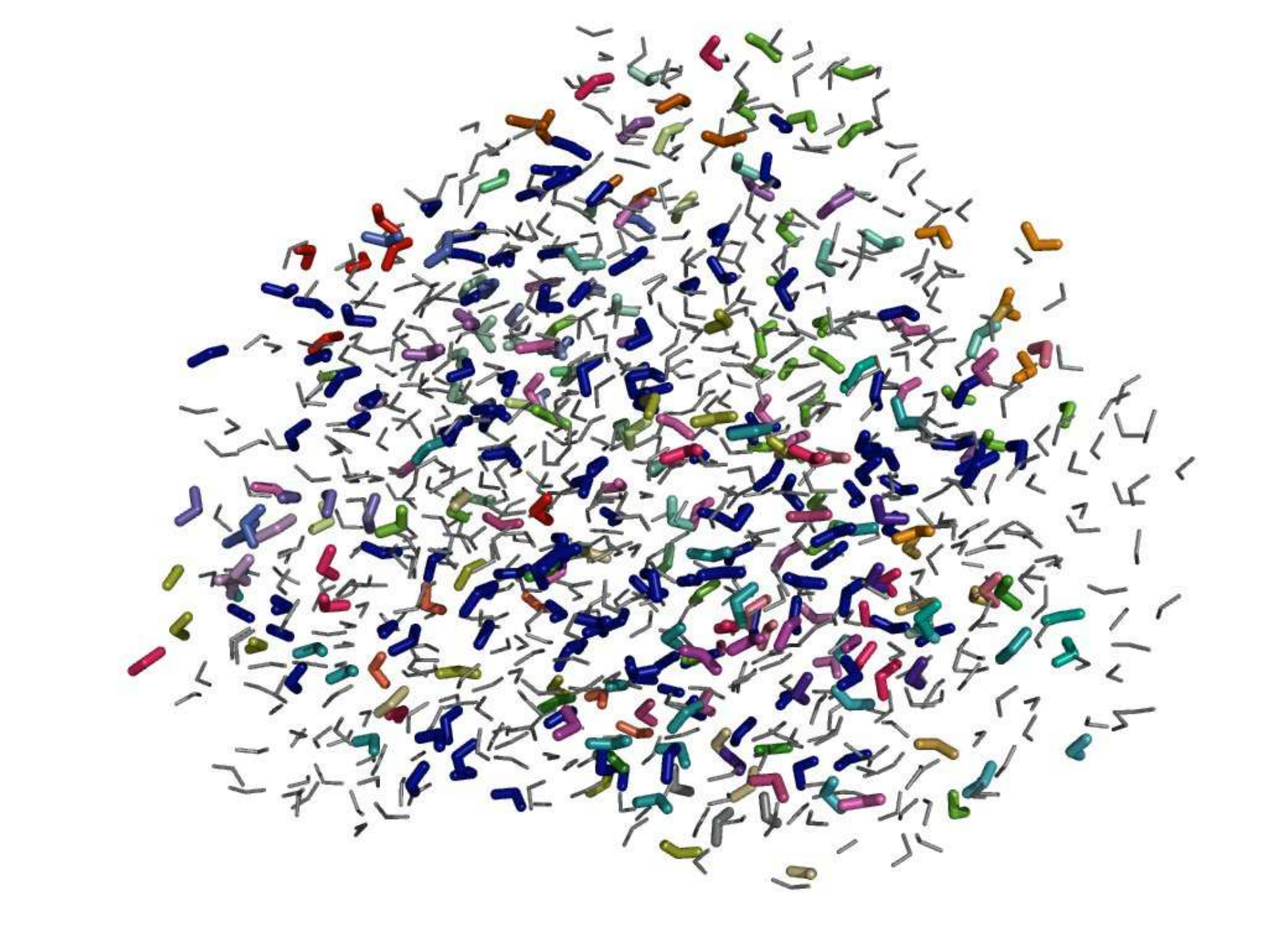}
\caption{A single snapshot of a simulated drop of 1000 water molecules, in vacuum. Coloring is by membership in individual co-moving clusters, as identified by \textit{TIMME} at a cutoff of $\sigma_{c} / a = $ 0.54 (the dashed line in Fig. \ref{fig:waterDropDilutionPlot}). Those co-moving clusters spanning more than one water molecule are shown as thick colored sticks, while those spanning only a single water molecule are shown as thin grey sticks. At this cutoff, the largest cluster spans 17 molecules. }
\label{fig:waterDropClusters}
\end{center}
\end{figure}

This combination of interactions creates a more complicated hierarchy of clusters with a large number of intrinsic time and length scales\cite{stillinger:1974os, dang:1997il}, as revealed by \textit{TIMME} analysis the molecular dynamics trajectory. There is sufficient disorder in the system that the fraction of water molecules in a cluster of size 10 molecules or more follows a sigmoidal curve, as can be seen in Fig. \ref{fig:waterDropLargestClusterStatistics}. Clusters of associated water molecules are visible as colored stalactites dangling from the core in the dilution plot shown in Fig. \ref{fig:waterDropDilutionPlot}. A representative snapshot of the water simulation is shown in Fig. \ref{fig:waterDropClusters}. 

\subsection{Proteins}

Proteins are linear polymers of amino acids. Their structure depends on their sequence, subject to the environment in which they are expressed\cite{eisenhaber:1995fu, gilmanshin:1997kb, rossmann:1981ye, mann:2003qf, spiro:2002qo}. A typical protein has a relatively rigid internal scaffolding that acts as a support to hold more flexible domains into specific relative positions and orientations. Fexible domains can serve to facilitate protein-protein interactions, to bind substrates, or to serve as a hinge between two or more relatively rigid scaffolds. 

Understanding which domains are more rigid and which are more flexible can give significant insight into how a protein performs its functions as a molecular machine. In addition to the intrinsic interest of knowing how a protein works, such understanding can be of practical use for suggesting possible targets for drug design to facilitate or inhibit a protein's performance \cite{drews:2000kl, ring:1993rw, chen:2001qa, teague:2003oz, klebe:2000mi}. 

A number of experimental techniques are used to determine protein structures. Two of the most common are X-ray diffraction \cite{kendrew:1960xr, kendrew:1958pi} and Nuclear Magnetic Resonance (NMR)\cite{braun:1981nx, wagner:1987fe}.

X-ray diffraction techniques diffract a beam of X-rays from highly purified crystals of a protein. Solution of the phases of the resulting diffraction pattern gives a reciprocal space representation of electron density in the protein. Computation of the inverse Fourier transform of that reciprocal space representation provides the real-space electron density, from which a representative structure for the protein can be inferred. 

NMR techniques measure internal distance and orientation constraints within an ensemble of structures. By imposing these constraints on a molecular dynamics simulation of the protein, an ensemble of structures consistent with the constraints can be generated. Assuming that a sufficiently large number of constraints have been measured, the structures reported from an NMR experiment will reflect some of the more populated states in the conformational space of the protein. 

Rigid clusters can be inferred from static structures using a graph theoretic structure analysis system, called `Floppy Inclusions and Rigid Substructure Topography', or \textit{FIRST}\cite{jacobs:1995wd, jacobs:2001nx}. \textit{FIRST} balances constraints and degrees of freedom to determine which regions of a molecule will be rigid and which will be flexible, under a specific set of constraints. The constraints used for this analysis include covalent bonds, hydrophobic tethers, and hydrogen bonds with strength greater than a user-specified cutoff. Alternatively, Molecular Dynamics simulations can be performed to explore which domains of a static protein structure are relatively rigid and which are relatively flexible. 

\textit{TIMME} was originally developed by us to aid in the analysis of ensembles of protein conformations, and to provide an alternative to \textit{FIRST} when more than a single conformation was available. Conceptually, \textit{FIRST} rigid clusters correspond roughly to \textit{TIMME} co-moving clusters at some cutoff. By selecting an appropriate cutoff, static structures, such as those reported from X-ray crystallography experiments, can be compared to ensembles of two or more structures, such as those reported from Nuclear Magnetic Resonance (NMR) structure determination experiments. 

When analyzing molecular frameworks, it is useful to consider only nearby pairs of bonded atoms to avoid identifying discontinuous clusters and to improve performance. Adjacent bonded atoms must be excluded because otherwise, $\sigma$ is will only be a measure of thermal fluctuations in bond lengths and angles. Also, if no thermal bond length and angle fluctuations were present in an ensemble, then any bonded molecule would be identified as a single co-moving cluster at any cutoff. Relative motion between third-nearest neighbor atoms reflects bond torsion, possibly restricted by distant steric and charge constraints. Even considering this, a well folded structure will have limited local motion and so the values of $\sigma$ will typically be small compared with the system.

\subsubsection{Barnase}

Barnase is a 110-residue water-soluble protein that is produced and secreted by \textit{Bacillus amyloliquefaciens}. By degrading RNA, barnase can stop protein synthesis and eventually kill neighboring cells. To protect itself from the action of barnase, \textit{B amyloliquefaciens} coexpresses an inhibitor called barnstar\cite{hartley:1989kl}. Within a cell expressing barnase, barnase is very tightly bound to barnstar\cite{schreiber:1995sp}. For this reason, barnase is commonly used in studies of protein-protein interactions, and protein dynamics. 

As an example of the correspondence between rigid clusters identified from static structures and those from dynamic structures, a \textit{FIRST} analysis was performed for a static X-ray structure of barnase (pdb ID: 1a2p\cite{martin:1982bs}), and compared with \textit{TIMME} analysis of a corresponding NMR ensemble (pdb ID: 1bnr\cite{bycroft:1991qf}), and of a 0.1 ns molecular dynamics simulation starting from the X-ray structure.

\begin{figure}[H]
\begin{center}
\includegraphics[width=\columnwidth]{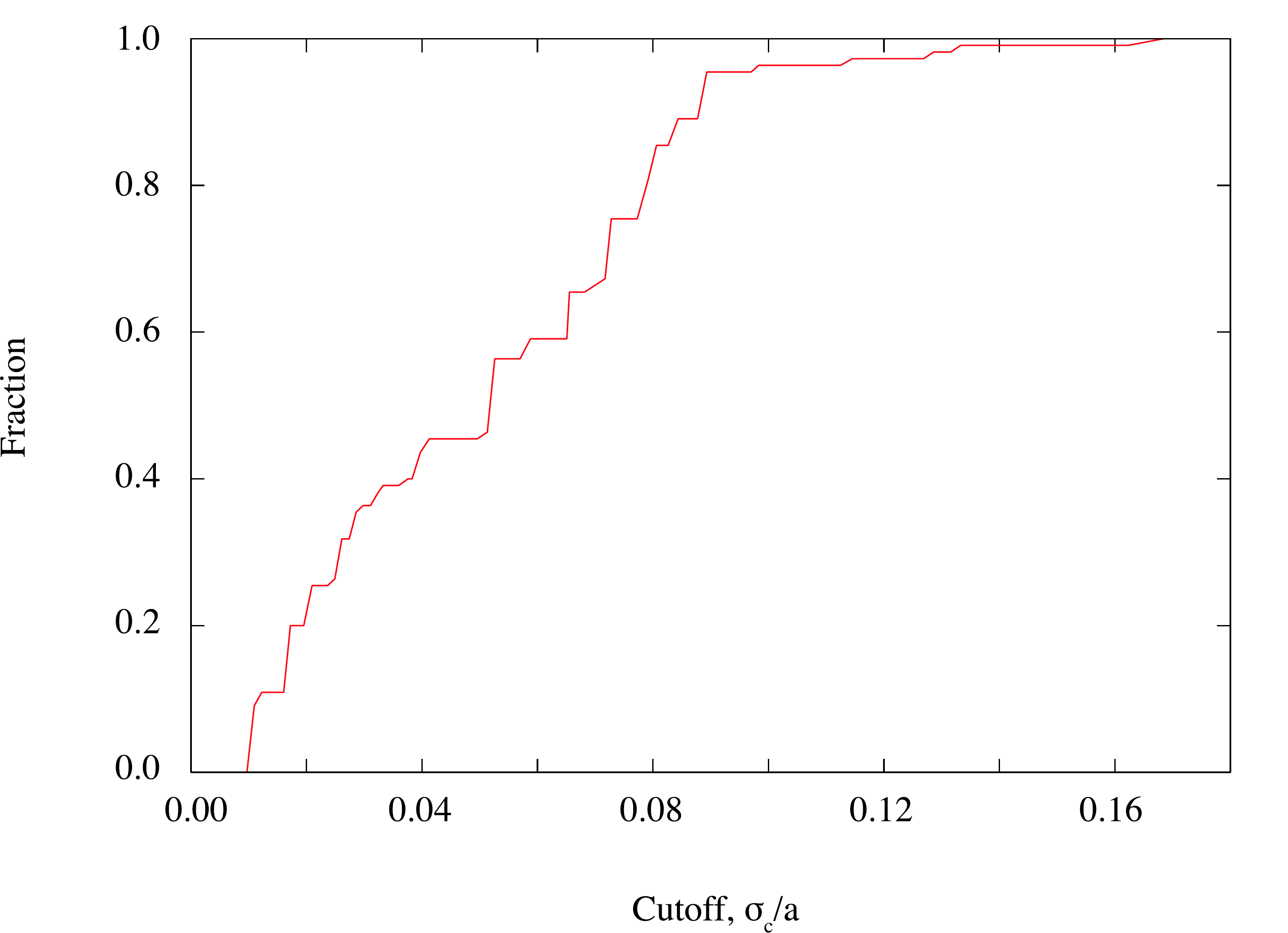}
\caption{The fraction of main-chain $\alpha-$carbon atoms in co-moving clusters of 10 or more main-chain  $\alpha-$carbon atoms in barnase (pdb ID: 1bnr\cite{bycroft:1991qf}) computed by \textit{TIMME}. The x-axis shows the cutoff, $\sigma_{c}$ scaled by the carbon-carbon bond length ($a = $ 1.54 \AA).}
\label{fig:barnaseFractionGreaterThan10}
\end{center}
\end{figure}

We found that the best agreement between \textit{FIRST} and \textit{TIMME} results occurred with respective cutoffs of approximately -0.05 kcal/mol and $\sigma_{c}/a = $ 0.04. Unlike the freely-jointed chain, water, and PMMA sphere systems, the fraction of atoms in the largest comoving cluster of barnase, shown in Fig. \ref{fig:barnaseFractionGreaterThan10}, is non-sigmoidal, reflecting greater structure imposed by the covalent bond network and non-covalent interactions. Although they are not identical, a casual inspection of Fig. \ref{fig:barnaseNMRandXrayComposite} reveals significant overlap between the rigid clusters identified by \textit{FIRST}, and the co-moving clusters identified by \textit{TIMME} for the NMR and a 0.1ns NAMD\cite{phillips:2005lr} MD ensembles. 

\begin{figure}[H]
\begin{center}
\includegraphics[width=\columnwidth]{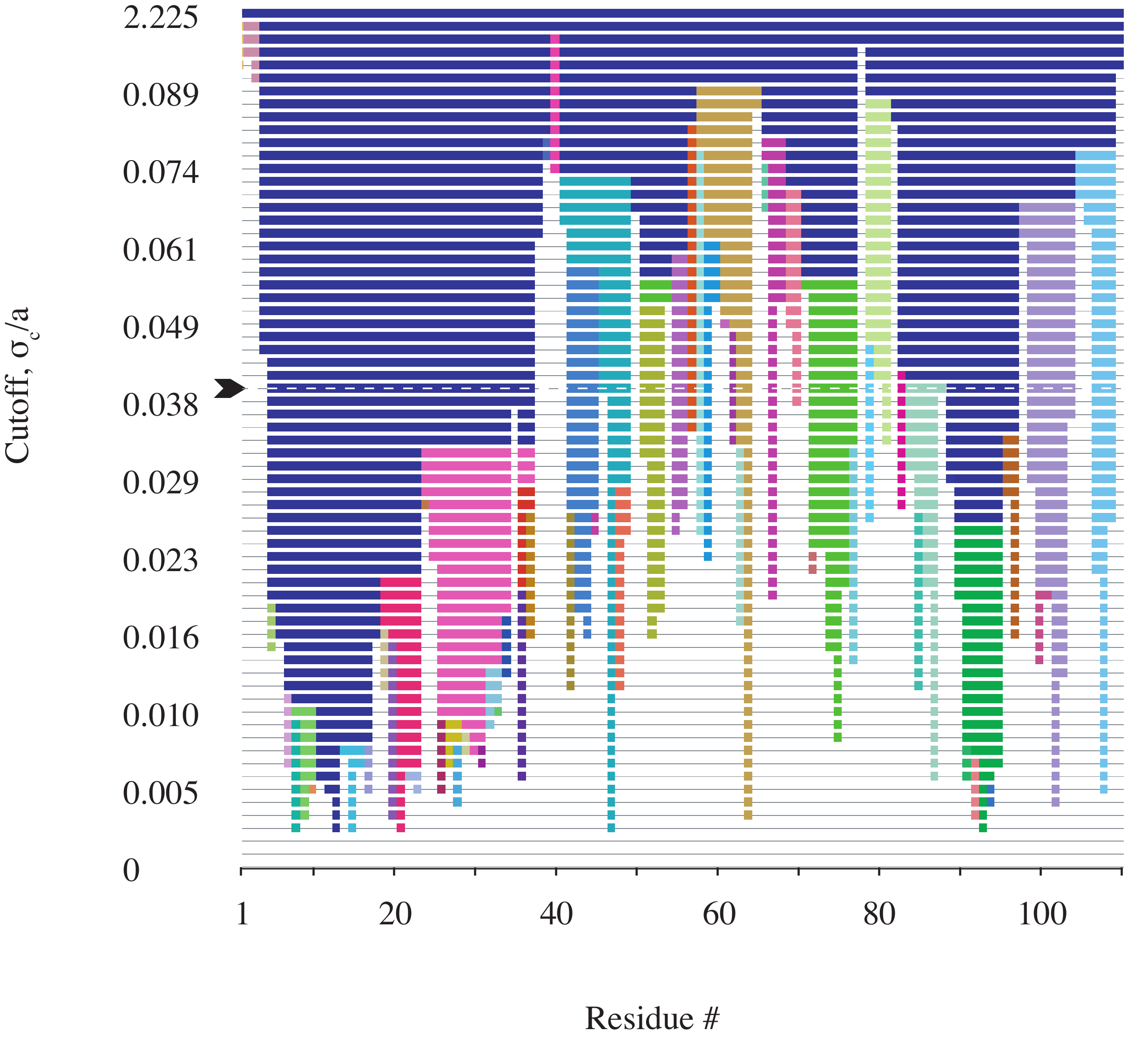}
\caption{Dilution of the clusters identified by \textit{TIMME} for the main-chain $\alpha-$carbon atoms, ordered by residue number, from an NMR ensemble of barnase (pdb ID: 1bnr\cite{bycroft:1991qf}). The side-chain atoms which are suppressed for clarity have higher $\sigma$ values and tend to join the clusters at higher $\sigma_{c}$. The y-axis shows the cutoff, $\sigma_{c}$ scaled by the carbon-carbon bond length ($a = $ 1.54 \AA). The horizontal dashed line shows the value of the cutoff used in Fig. \ref{fig:barnaseNMRandXrayComposite}}.
\label{fig:1bnrDilutionPlot}
\end{center}
\end{figure}
Proteins have an intrinsic primary sequence, which provide a natural site label for dilution plots, such as that shown in Fig. \ref{fig:1bnrDilutionPlot}. With appropriate choice of cutoff, the rigid clusters identified by \textit{FIRST} for a static X-ray structure (pdb ID: 1bnr\cite{bycroft:1991qf}) corresponds well with those identified by \textit{TIMME} for an NMR ensemble (pdb ID: 1bnr\cite{bycroft:1991qf}) and a 0.1 ns \textit{NAMD}\cite{phillips:2005lr} molecular dynamics simulation of the X-ray structure. This correspondance can be seen in Fig. \ref{fig:barnaseNMRandXrayComposite}
\begin{figure*}[!ht]
\begin{center}
\includegraphics[width=6 in]{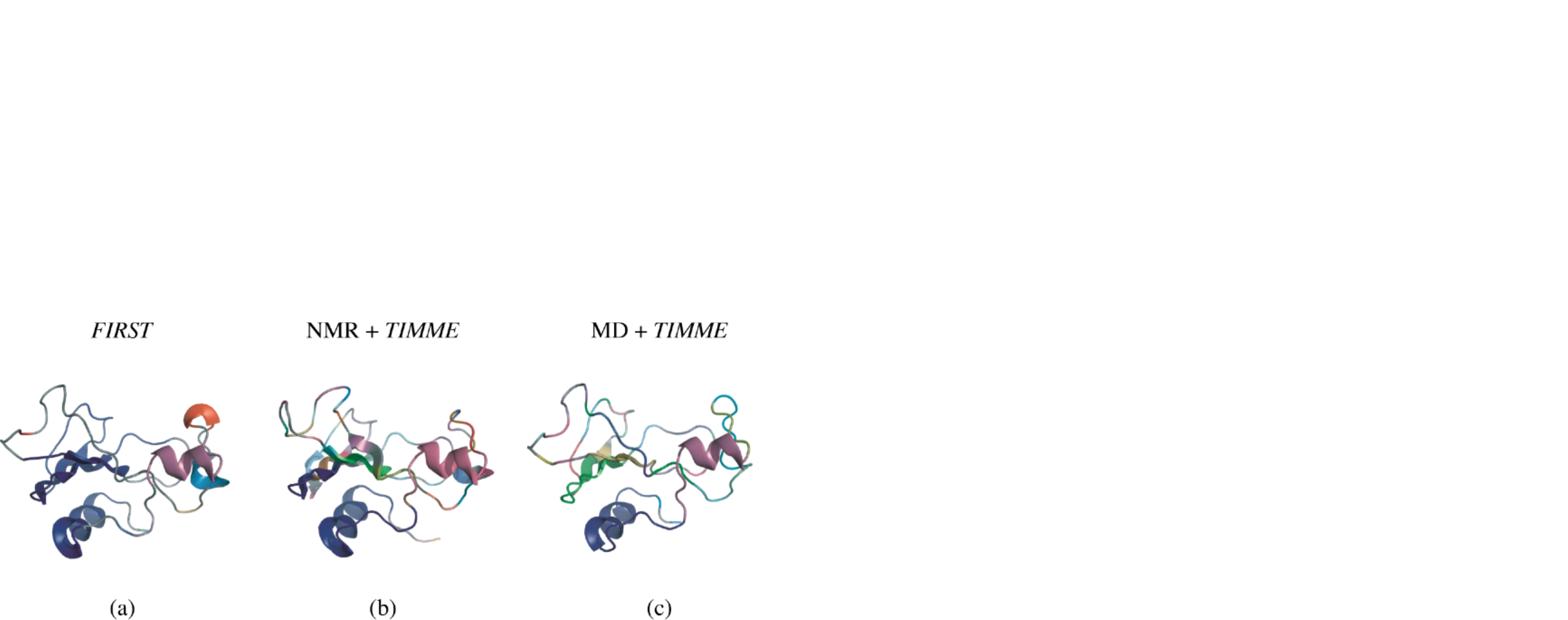}
\caption{\label{fig:barnaseNMRandXrayComposite}Analysis of the bacterial ribonuclease, barnase, by \textit{FIRST} and \textit{TIMME}. Panel (A) shows a rigid cluster decomposition of an X-ray crystallography structure (pdb ID: 1a2p, chain A\cite{martin:1982bs}), at an energy cutoff of -0.048 kcal/mol. 
Individual rigid clusters are colored in order of their size, with the largest cluster, or `rigid core', in dark blue. Panel (b) shows a co-moving cluster decomposition of an NMR ensemble in aqueous solution (pdb ID: 1bnr\cite{bycroft:1991qf}), at $\sigma_{c}/a = $ 0.04 (dashed line in Fig. \ref{fig:1bnrDilutionPlot}) applied to a single snapshot for visualization. As in panel (a), individual co-moving clusters are colored to match the corresponding clusters in panel (a). Panel (c) shows a co-moving cluster decomposition of an MD simulation based on an X-ray crystallography structure (pdb ID: 1a2p, chain A\cite{martin:1982bs}), at a cutoff, $\sigma_{c}/a = $ 0.068, with clusters colored to match corresponding clusters in panel (a). }
\end{center}
\end{figure*}

\subsubsection{Adenylate Kinase}

Adenylate Kinase (ADK) is a ubiquitous enzyme thatcatalyzes formation of Adenylate TriPhosphate (ATP) and Adenylate MonoPhosphate (AMP) from two molecules of Adenylate DiPhosphate (ADP), as well as the reverse reaction, 2 ADP $\longleftrightarrow$ AMP + ATP. 

ATP is a nucleotide -- one of the bases incorporated into DNA and RNA. It is also a common cellular energy currency. It is used by many different proteins to drive endothermic reactions. Energy is stored by the bonding of phosphate to AMP or ADP in a dehydration reaction. By coupling hydrolysis of ATP to ADP to an endothermic reaction, enzymes within biological cells are able to drive energetically unfavorable reactions\cite{astumian:1996ij, cline:1992th, cross:1982bs}. In addition to the role in cellular energetics, phosphate groups from ATP are used in regulation. Enzymes called protein kinases cleave phosphate from ATP and covalently bond the phosphate to proteins. Phosphorylation acts as a switch by altering the activity of a protein. A common example involves transduction by cascades of protein kinases of a signal from a cellular sensor to the nucleus of a cell in order to up- or down-regulate expression of one or more proteins \cite{hardie:246fy, kyriakis:1996fv, huwiler:1994la, zhou:1995dz}. 

ADK ensures that the concentrations of ATP, ADP and AMP within a cell quickly reach equilibrium, to facilitate energy production and use\cite{colowick:1943hc,gerstein:1993dp,miron:2004uq}.

\begin{figure}[H]
\begin{center}
\includegraphics[width=\columnwidth]{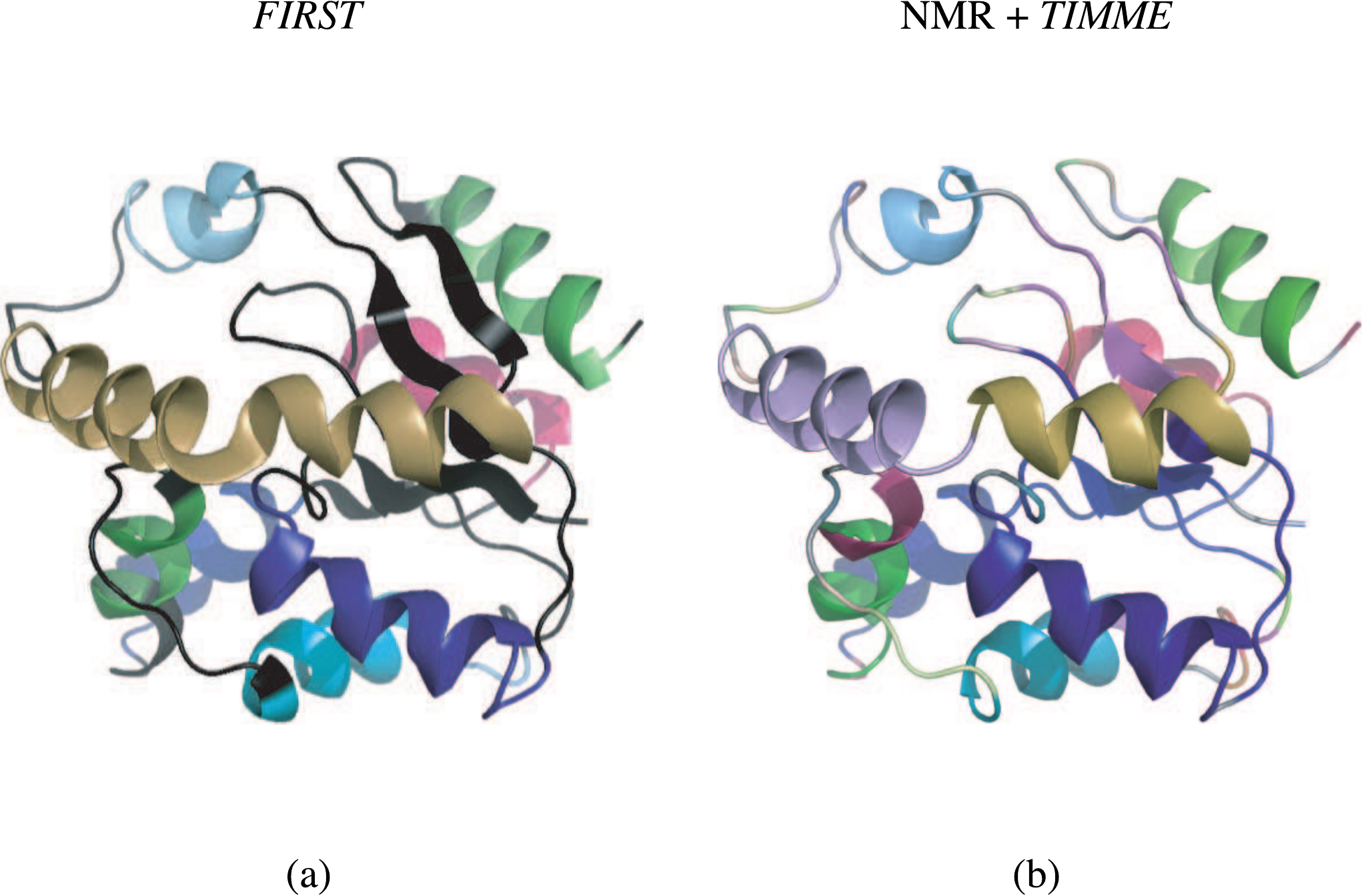}
\caption{\label{fig:adkNMRandXrayComposite}Analysis of Adenylate Kinase by \textit{FIRST} and \textit{TIMME}. Panel (a) shows a rigid cluster decomposition of a model from an NMR ensemble in aqueous solution (pdb ID: 1p4s \cite{miron:2004uq}), at a hydrogen-bond energy cutoff of -0.185 kcal/mol.
Individual rigid clusters are colored in order of their size, with the largest cluster, or the `rigid core', in dark blue. Panel (b) shows a co-moving cluster decomposition of an NMR ensemble of ADK in aqueous solution (pdb ID: 1p4s \cite{miron:2004uq}), at $\sigma_{c}/a = $ 0.09, applied to a single snapshot for visualization. As in panel (a), individual co-moving clusters are colored according to their size in the leftmost figure, with the largest co-moving cluster, or the `co-moving core', colored in dark blue. }
\label{fig:adkTIMMErcd}
\end{center}
\end{figure}

\textit{FIRST} analysis was performed for a static snapshot from an NMR ensemble for ADK (pdb ID: 1p4s model: 1 \cite{miron:2004uq}), and compared with \textit{TIMME} analysis of the complete NMR ensemble (pdb ID: 1p4s\cite{miron:2004uq}). Cutoff distances were selected to maximize the overlap between \textit{FIRST} rigid clusters and \textit{TIMME} co-moving clusters. \textit{FIRST} analysis was performed with hydrogen bonds stronger than -0.185 kcal/mol included in the constraint counting. \textit{TIMME} analysis was performed with $\sigma_{c}/a = $ 0.09, where $a$ is the carbon-carbon bond length, 1.54 \AA. At these cutoffs, there is significant overlap, as can be seen in Fig. \ref{fig:adkTIMMErcd}.

\section{Discussion}

Understanding and documenting the evolution of the particles in inhomogenous many-particle systems is the subject of this article. Computer simulations and experimental observation of these large systems go hand-in-hand in studying these large systems. The results of such simulations or experiments include such a large quantity of information that data mining techniques are essential for understanding and revealing the underlying properties of such systems. A number of approaches have been developed to that end, including factor analytic and clustering techniques.

A new method for identifying clusters of sites that move in concert is introduced in this paper. This approach, implemented as the Tool for Identification of Mobility in Macromolecular Ensembles (\textit{TIMME}), combines a number of the strengths of factor analysis and hierarchical classification. 

The \textit{TIMME} algorithm has been applied to a number of sample test cases. First, the method was demonstrated on a freely-jointed polymer chain with no other intrinsic structure, where an analytic solution was available for comarison. Second, PMMA spheres were analyzed using data obtained via confocal microscopy, Third, the technique was applied to a Molecular Dynamics simulation of a drop of water, which exhibited a complicated hierarchy of organization due to hydrogen bonding and other weak interactions between different adjacent water molecules. Finally, two proteins in aqueous solution observed by NMR structure determination experiments have been analyzed. In both protein cases, the observed hierarchical structure we compute corresponds to known properties of the proteins, and also to the hierarchical structure determined from static X-ray structures with the software \textit{FIRST}.
 
Incomplete sampling is arguably the most severe limitation of the \textit{TIMME} algorithm. Because the algorithm requires an estimate of $\sigma_{a b}$ for each pair of sites $a$ and $b$, it is sensitive to under-sampling, bias and errors. This limitation is not unique to \textit{TIMME}, but it is a fundamental problem for any analysis approach based on incomplete information\cite{clarage:1995qf}.

One complementary approach to analyzing dynamics of many-bodied systems is to perform a Principal Components Analysis (PCA) on a representative collection of $3n$ dimensional snapshot states. PCA is a linear transform that identifies the specific directions of motion within the system that contribute most to the system's overall motion. A PCA can be thought of as similar to a Fourier transform, except that the basis vectors are empirically derived to satisfy certain properties. In particular, the projection of all snapshot states onto the first PCA basis vector has the largest variance of any linear projection; with the projection onto the second PCA basis vector having the largest variance of any other linear projection orthogonal to the first, and so on.

There are a number of algorithms for performing a PCA. One common approach begins by computing a covariance matrix for the $3n$ dimensional snapshot states. If $\underline{x}_a$ represents the coordinate vector of $a$ in a basis, and $\underline{x}_b$ the coordinates of $b$, then the covariance of $a$ and $b$ is  
 \begin{eqnarray}
cov_{a b} &=& \langle\left(\underline{x}_a - \langle\underline{x}_a\rangle \right) \cdot \left(\underline{x}_b  - \langle\underline{x}_b\rangle \right) \rangle
\end{eqnarray}
where the expectation value $\langle \underline{x}\rangle$ is shorthand for the arithmetic mean value of $\underline{x}$. 

The covariance of $a$ and $b$ indicates how much $a$ and $b$ vary together. If the positions of $a$ and $b$ are correlated, then $cov_{a b} > 0$. If they are anti-correlated, then $cov_{a b} < 0$, and if they are uncorrelated, then $cov_{a b} = 0$. In this way, $cov_{a b}$ can be used to identify the correlated motions within an ensemble. 

The bilinear, symmetric $3N \times 3N$ matrix $cov_{a b}$ is independent of uniform global translations and rotations of all snapshots, but not independent of relative translations and rotations between snapshots. 

Diagonalizing the covariance matrix yields a set of basis vectors for the system. Sorted in decreasing order of the corresponding diagonal elements, these vectors correspond to the directions of motion with decreasingly large contribution to the variance. These basis vectors are called principal components and hence the approach is called a Principal Components Analysis (PCA)\cite{wold:1976uq, hotelling:1933kx, balsera:1996bh}. 

Most often, this process is implemented by performing a Singular Value Decomposition (SVD) on the covariance matrix. This matrix factorization can roughly be thought of as a generalized eigenvector/eigenvalue decomposition. In an SVD, a $m \times n$ matrix $M$ is factorized in the form
\begin{equation}
\underline{\underline{M}} = \underline{\underline{U}}\  \underline{\underline{\Sigma}}\  \underline{\underline{V}}^{\dag}
\end{equation}
where $\underline{\underline{U}}$ is an $m \times n$ matrix of basis vectors for the output range, $\underline{\underline{\Sigma}}$ is an $n \times n$ matrix with the singular values as the only nonzero entries down the diagonal, and $\underline{\underline{V}}^{\dag}$ is the conjugate transpose of an $n \times n$ matrix of basis vectors for the input domain\cite{press:1992ve}.

Decomposition into co-moving clusters in \textit{TIMME} at a given cutoff corresponds roughly to the model-free approach of decomposing NMA or PCA covariance matrices into a block-diagonal form\cite{hayward:1997lq}. The individual blocks in the matrix correspond to distinct quasi-rigid domains, which can be thought of as co-moving clusters. In model-free methods, a fixed set of co-moving clusters are chosen initially, and used throughout the analysis. \textit{TIMME} extends this approach to decomposing an entire system into successively smaller quasi-rigid domains ranging from the entire system as a whole to the smallest indivisible rigid building blocks. 

\section{Conclusions}

The algorithm \textit{TIMME} developed in this paper, is an objective and systematic approach to analyzing ensembles of snapshots of a dynamical system. Co-moving cluster decompositions based on pair-distance statistics provide a useful way to understand detailed features in such systems that are not readily visible to an unaided observer. To do this, it is necessary to have two or more snapshots of particle configurations are available for analysis.

Because the algorithm relies on internal pair distances, the results are independent of choice of coordinates, and system-wide translations and rotations between snapshot configurations. The algorithm makes no assumptions about the number or type of objects being analyzed, and can be applied to any system with two or more snapshot realizations with a metric that allows a pair distance to be measured and monitored. Although all of the examples shown here were embedded in 3-dimensional space and modeled physical objects, this method could be applied 2-dimensional and also to high-dimensional abstract data sets to identify a hierarchy of clusters. 

In order to apply the algorithm, individual particles must be enumerated so that they can be tracked. This can be accomplished in two ways. In a protein, the atoms are naturally numbered by their position along the polypeptide chain. In a collection of identical objects like PMMA spheres, the situation is more complex, and the snapshots must be taken at sufficiently short time intervals such that no sphere has moved further than about a radius so that it can be continuously tracked unambiguously.

Finally we note that rigid cluster do not have to be contiguous in three and higher dimensions \cite{whiteley:2005lr, chubynsky:2007lr}. This is a very counter-intuitive situation where mutually rigid and con-contiguous pieces can form a single rigid cluster, while being embedded in a fluid. The present algorithm could be used to identify such (comparatively rare) situations, by monitoring all pairs and not just close neighbors, in order to build up the rigid clusters. To our knowledge, no such situations have ever been identified in the laboratory.

\begin{acknowledgments}
We thank M. Chubynsky, Dan Farrell, Brandon Hespenheide, Craig Jolley, and Stephen Wells for useful discussions, and Yongxiang Gao for providing the raw data of the confocal microscopic observations of PMMA spheres. We acknowledge support from the NIH via grant 1 R01 GM 67249-01 and from NSF grants DMR-0304391 and DMR-0078361. \end{acknowledgments}

\appendix

\section{Program Availability}
\textit{TIMME} has been implemented as a module of \textit{FIRST}. The combination \textit{FIRST/TIMME} is available gratis to academic users via FlexWeb (http://flexweb.asu.edu/software/timme). The source code can be downloaded as part of the \textit{FIRST} suite or the program can be run interactively online using FlexWeb. Note that the graphics used in Figs. (\ref{fig:PMMASpheresKissingNetwork}, \ref{fig:waterDropClusters}, \ref{fig:barnaseNMRandXrayComposite}, and \ref{fig:adkNMRandXrayComposite}) were generated with PyMol \cite{delano:2002rr}. 

\bibliography{hierarchialRigidityFromPairDistanceFluctuations}

\end{document}